\documentclass[10pt,letterpaper]{article}
\usepackage[top=0.85in,left=1in,footskip=0.75in,marginparwidth=2in]{geometry}

\usepackage[utf8]{inputenc}

\usepackage{cite}
\usepackage{amsmath}
\usepackage{gensymb}
\usepackage{xfrac}
\usepackage{soul}
\usepackage{xcolor}

\usepackage{nameref}
\usepackage[hidelinks]{hyperref}

\usepackage{microtype}
\DisableLigatures[f]{encoding = *, family = * }

\usepackage{changepage}

\usepackage[aboveskip=1pt,labelfont=bf,labelsep=period,singlelinecheck=off]{caption}

\makeatletter
\renewcommand{\@biblabel}[1]{\quad#1.}
\makeatother

\usepackage{lastpage,fancyhdr,graphicx}
\usepackage{epstopdf}
\fancyheadoffset[L]{2.25in}
\fancyfootoffset[L]{2.25in}

\usepackage{color}

\definecolor{Gray}{gray}{.25}

\usepackage{graphicx}

\usepackage{sidecap}

\usepackage{wrapfig}
\usepackage[pscoord]{eso-pic}
\usepackage[fulladjust]{marginnote}
\reversemarginpar

\begin{document}
\vspace*{0.35in}

% title goes here:
\begin{center}
{\Large
\textbf\newline{Effect of substrate temperature on the optoelectronic properties of DC magnetron sputtered copper oxide films}
}
\newline
% authors go here:
\\
Aarju Mathew Koshy\textsuperscript{1},
A. Sudha\textsuperscript{1, 2},
Satyesh Kumar Yadav\textsuperscript{3, 4},
Parasuraman Swaminathan\textsuperscript{1,2*}

\bigskip
\textsuperscript{1}Electronic Materials and Thin Films Lab,
Dept. of Metallurgical and Materials Engineering, \\
Indian Institute of Technology, Madras, Chennai, India \\
\textsuperscript{2}Ceramics Technologies Group - Center of Excellence in Materials and \\Manufacturing for Futuristic Mobility,
Dept. of Metallurgical and Materials Engineering, \\
Indian Institute of Technology, Madras, Chennai, India
\\
\textsuperscript{3}Materials Design Group, Department of Metallurgical and Materials Engineering, \\
Indian Institute of Technology, Madras, Chennai, India \\
\textsuperscript{4}Center for Atomistic Modelling and Materials Design,\\ Indian Institute of Technology, Madras, Chennai, India
%\bf{2} Affiliation B

\bigskip
*Email: swamnthn@iitm.ac.in

\end{center}

\section*{Abstract}
Copper oxide thin films are deposited on quartz substrates by DC magnetron sputtering and the effect of deposition temperature on their optoelectronic properties is examined in detail. Scanning Electron Microscopy (SEM), X-ray diffraction (XRD) analysis, Raman spectroscopy, UV-Vis spectroscopy, and four-probe sheet resistance measurements are used to characterize the surface morphology, structural, optical, and electrical properties respectively. Deposition is carried out at room temperature and between 200 and 300 \textdegree C. XRD analysis indicates that the oxide formed is primarily Cu$_2$O and the absorption spectra show the films have a critical absorption edge at around 300 nm. The sheet resistance gradually decreases with increase in deposition temperature thereby increasing the conductivity of these thin films. Also observed is the increase in band gap from 2.20 eV for room temperature deposition to 2.35 eV at 300 \textdegree C. The optical band gap and the variation of sheet resistance with temperature shows that the microstructure plays a vital role in their behavior. These transformation characteristics are of huge technological importance having variety of applications including transparent solar cell fabrication.

\bigskip

\noindent \textbf{Keywords:} Copper oxide; Reactive sputtering; Optoelectronic properties; Substrate temperature

\bigskip
% now start line numbers
%\linenumbers
\newpage
% the * after section prevents numbering
\section{Introduction}
Copper oxide is a $p$-type nontoxic metal oxide semiconductor, possessing a wide range of applications in gas sensors, as hole transport layers in solar cells, and also in electrochromic devices \cite{wong2016current,richardson2001electrochromism}. There are normally two oxides of copper: cuprous or copper (I) oxide (Cu$_2$O), with a direct band gap slightly above 2.0 eV and cupric or copper (II) oxide (CuO) having an indirect band gap between 1.3-2 eV \cite{nolan2006p,kunti2016comparative,rydosz2018use,aktar2020solution}. A metastable, intermediate copper oxide compound between the above mentioned two phases, Cu$_4$O$_3$, having catalytic properties has also been reported \cite{debbichi2012vibrational,murali2018thermal}. Cu$_2$O crystallizes in a cubic structure with the lattice parameter a = 0.427 nm, CuO and Cu$_4$O$_3$ have monoclinic and tetragonal structures respectively. Cu$_2$O is the more stable oxide at low temperature while CuO is stable at high temperature \cite{zheng2018phase}. Due to its tunable band gap, the copper (I) oxide acts as a good absorber in polycrystalline hetero junction solar cells, e.g., ZnO/Cu$_2$O, In$_2$O$_3$/Cu$_2$O, ITO/Cu$_2$O and ZnIn$_2$O$_4$/Cu$_2$O \cite{wong2016current,winkler2018solution,johan2011annealing}. The $p$-type semiconducting behavior of Cu$_2$O arises due to its non-stoichiometry \cite{bose2005electrical}. The high optical absorption coefficient in the visible range and electrical properties make the material highly useful for optoelectronic applications, such as a channel material for $p$-type oxide thin film transistors \cite{papadimitropoulos2005deposition,paredes2018ultrasonic}. Generally, the oxide properties depend on the growth method employed, deposition conditions, and post-annealing processes (if any) \cite{chen2016effect}. To date, the oxides of copper have been prepared using several deposition techniques that include thermal evaporation, sol-gel method, chemical vapor deposition, and magnetron sputtering \cite{al2019effect,babar2015surface,raship2017effect,zheng2018phase,low2013morphology}. Among the various techniques, DC magnetron sputtering is a versatile method to prepare copper oxide films, as the sputtering of the metallic target in the presence of various reactive gas environments provides large area deposition, with good uniformity and adhesion between the substrate and thin films and good repeatability. Although several studies are reported in the literature \cite{zhang2015cuo,papadimitropoulos2005deposition,wang1996electronic} on the oxides of copper, this work is an attempt to prepare copper oxide films from metallic copper by DC reactive magnetron sputtering in the presence of a mixture of argon and oxygen gases. The effect of deposition temperature on the structure, optical, and electrical properties of the deposited films, are extensively studied. The stabilization of the copper (I) oxide phase and the change in its properties with substrate temperature will be useful in device fabrication applications.

\section{Experimental Section}
Copper oxide thin films of approximately 70 nm were grown on quartz substrates using DC magnetron reactive sputtering using a pure Cu target. The substrates (with dimensions 2 × 2 cm$^2$) were cleaned ultrasonically in acetone/ethanol/deionized water for 15 min each (in this order) and then dried in air before loading in the sputtering chamber. The background pressure of the sputtering chamber was set at $5 \times 10^{-6}$ mbar, and the working pressure was recorded at $1.2 \times 10^{-3}$ mbar. Argon gas of purity 99.999$\%$ was used as the sputtering gas filling the chamber along with oxygen gas of purity 99.999$\%$ as the reactive gas. The substrate temperature was stabilized for 5 min prior before process. DC power of 90 W was used and deposition was done for 15 min keeping both the Ar and O$_2$ flow rate constant at 20 standard cubic centimeters per minute (sccm). Deposition was done at room temperature and at higher temperatures, namely 200, 250, and 300 \textdegree C.

Structural characterization was carried out using RIGAKU Miniflex 600 Benchtop 6th Generation X-ray diffractometer with Cu-K$\alpha$ radiation of wavelength 0.154 nm in grazing angle. The surface morphology was imaged using scanning electron microscopy (SEM) (FEI Inspect F). The transmittance of the films was recorded using an UV-Vis spectrophotometer (Jasco V-570). The electrical properties of films were measured using a four-point probe instrument (Jandel RM3000). Raman spectroscopy was done using the Alpha300 R advanced Raman imaging system.

\section{Results and discussions}
\subsection{Structural properties}

\begin{figure}[h]
\centering
  \includegraphics[width=8.3cm]{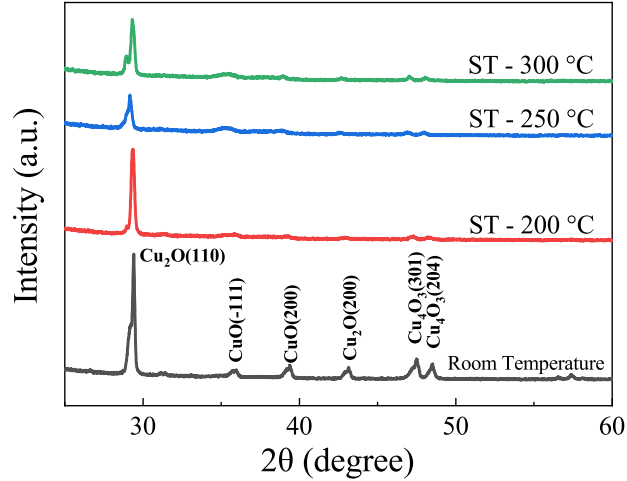}
  \caption{The XRD patterns of copper oxides films deposited at different substrate temperatures (referred as ST), ranging from room temperature to 300 \textdegree C. The predominant phase is Cu$_2$O.}  
  \label{fgr:fig1}
\end{figure}

XRD was used to understand the evolution of the phases as a function of the deposition temperature. Figure \ref{fgr:fig1} shows the XRD patterns of the films deposited at room temperature and higher substrate temperatures. From the XRD pattern, it is clear that the copper oxide film deposited at room temperature exhibits a mixed character with phases such as Cu$_2$O (ICDD number: 00-005-0667), CuO (ICDD number: 01-074-1021), and the metastable state Cu$_4$O$_3$ (ICDD number: 01-083-1665). No Cu peaks were observed in the XRD plot indicating that oxide formation is complete under these conditions. From the plot, well defined reflection peaks were observed at 2$\theta$ values of 29.4 (Cu$_2$O (110)), 35.9 (CuO (-111)), 39.2 (CuO (200)), 42.9 (Cu$_2$O (200)), 47.2 (Cu$_4$O$_3$(301)) and 48.3 (Cu$_4$O$_3$(204)) respectively. A preferred orientation of (110) is also observed for the Cu$_2$O phase. From this, it is evident that these films are polycrystalline in nature and the different phases formed correspond to the limited diffusivity and the formation of metastable phases under low temperature deposition. As the substrate temperature is increased, the intensity of the other phases except the Cu$_2$O phase starts decreasing, confirming the transformation into a single phase. However, for the substrate temperature of 300 \textdegree C, appearance of a weak peak or the peak splitting indicates a transformation to a lower symmetry phase. The appearance of sharp and strong peaks revealed overall good crystallinity of copper oxide thin films.

The data from the XRD plots are extracted and presented in table \ref{tbl:table1}, which consolidates the structural properties of the deposited films. The average grain size and the lattice strain are also estimated \cite{hasabeldaim2017effect,naveena2019comparative}. Using the prominent diffraction peak, corresponding to the (110) plane of Cu$_2$O, in each of the XRD patterns, the average grain size was estimated using the Debye-Scherrer formula given below.
 \begin{equation}  
D_{av} = \frac{K\lambda} {\beta _{hkl} cos\theta } 
\end{equation} 
where $K$ is taken as 0.9, $\lambda$ is the X-ray wavelength, $\theta$ is the diffraction angle, and $\beta_{hkl}$ the full width at half maximum (FWHM) of the corresponding diffraction peak (in radians). The lattice strain developed in the film is estimated by Willamson-Smallman (W-S) formula as given below
\begin{equation}
    \epsilon=\frac{\beta_{hkl}}{4 tan\theta}
\end{equation}

\begin{table*}
\small
  \caption{\ The structural properties of the copper oxide thin films deposited at different temperatures. The grain size and lattice strains are estimated using the Debye-Scherrer and Williamson-Smallman formula.}
  \label{tbl:table1}
  \begin{tabular*}{\textwidth}{@{\extracolsep{\fill}}llll}
    \hline
    Substrate Temperature (\textdegree C) & FWHM ($\beta_{hkl}$) & Grain size (nm) & Lattice strain ($\times 10^{-3}$) \\\hline
    Room Temperature & 0.098 $\pm$ 0.009 & 87.6 $\pm$ 7.6 & 1.6\\
    200 & 0.098 $\pm$ 0.005 & 71.5 $\pm$ 3.4 & 2.0 \\
    250 & 0.157 $\pm$ 0.010 & 87.5 $\pm$ 7.7 & 1.6\\
    300 & 0.079 $\pm$ 0.007 & 108.6 $\pm$ 11.3 & 1.3\\
    \hline
  \end{tabular*}
\end{table*}

It is clear from Table \ref{tbl:table1} that the grain size is small at lower deposition temperatures, as the deposited atoms instead of diffusing and integrating to the neighboring grains and raising their size condense and remain stuck to their regions forming new nuclei and clusters. As the deposition temperature increases, the grain size increases with a size of approximately 108.64 nm at 300 \textdegree C. This may be attributed to the increased mobility of the arriving atoms. Corresponding to the increased deposition temperature there is also a slight decrease in the lattice strain. The cause of internal strain might be due to the difference between thermal expansion coefficients of copper and the quartz substrate that is developed because of the displacements of atoms with respect to their reference lattice positions. Increased mobility with deposition temperature and a corresponding increase in average size decreases this lattice strain.

\begin{figure}[h]
\centering
  \includegraphics[width=8.3cm]{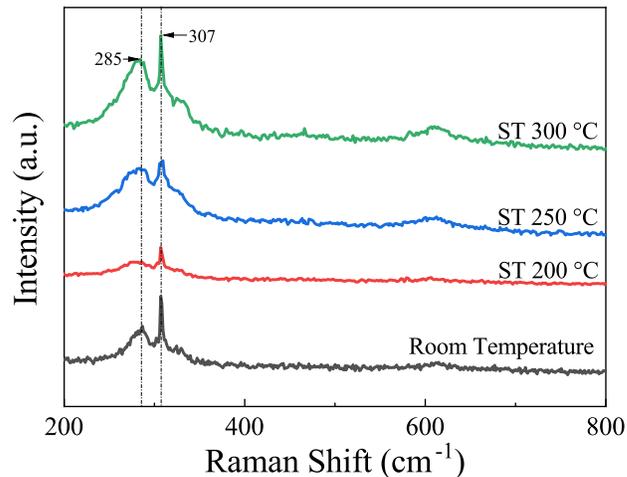}
  \caption{Raman spectra of copper oxide films deposited at different substrate temperatures. The predominant peaks can be indexed to Cu$_2$O and CuO.}
  \label{fgr:fig2}
\end{figure}

Figure \ref{fgr:fig2} plot the Raman spectra recorded for the copper oxide films deposited at different substrate temperatures. The Raman spectra show the characteristic phonon frequency of crystalline Cu$_2$O. It can be seen from the figures that there are two Raman peaks at 285 and 307 cm$^{-1}$. It is well established that the Raman spectra of Cu$_2$O has a peak at (308 cm$^{-1}$) \cite{solache2008raman}. Another weak peak observed near 285 cm$^{-1}$ is close to the peak normally attributed to the CuO phase \cite{mary2019sugarcane,chand2014structural}. With increase in deposition temperature, the spectra show an increase in this CuO peak compared to Cu$_2$O, which is indicative of a phase conversion with annealing. Although in the XRD patterns, CuO phase is almost absent (see figure \ref{fgr:fig1}), one can conclude from the Raman data that while the predominant phase formed is of the Cu$_2$O phase there is also some small amount of the CuO. Given that Raman spectroscopy is more surface sensitive than XRD\cite{akgul2014influence} it is possible that this phase is located closer to the film surface.

\subsection{SEM Analysis}
\begin{figure*}
 \centering
 \includegraphics[width=15.3cm]{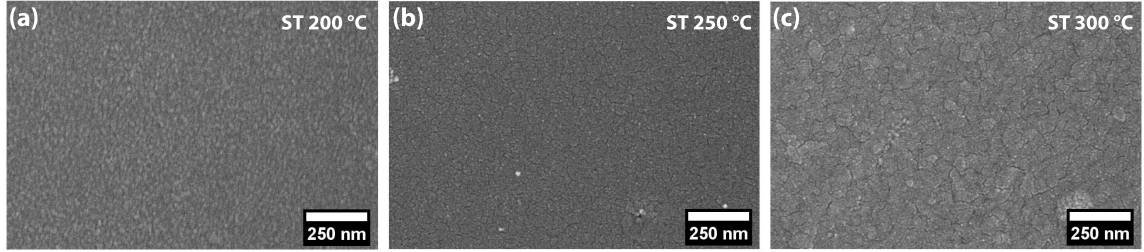}
 \caption{Scanning Electron Micrographs of the oxide films deposited at different substrate temperatures. The grain size increases with deposition temperature.}
 \label{fgr:fig3}
\end{figure*}

Figure \ref{fgr:fig3} shows the SEM images of the samples deposited at different substrate temperatures. It can be observed from the SEM images that the morphology of the films are significantly influenced by the change in deposition temperature during sputtering.The copper oxide film prepared at the substrate temperature of 200 \textdegree C exhibits spherical uniformly distributed grains. A similar behaviour is also observed at 250 \textdegree C. Also the grains are distributed uniformly with no porosity or agglomeration. For the film grown at 300 \textdegree C, the grain size has increased due to increased coalescence at higher deposition temperature \cite{gommes2019ostwald}. The results are also consistent with the crystallite size measurements from XRD (table \ref{tbl:table1}), which shows that there is nearly a 25\% increase in crystallite size going from 250 to 300 \textdegree C.  

\subsection{Optical Analysis}
\begin{figure*}
 \centering
 \includegraphics[width=15.3cm]{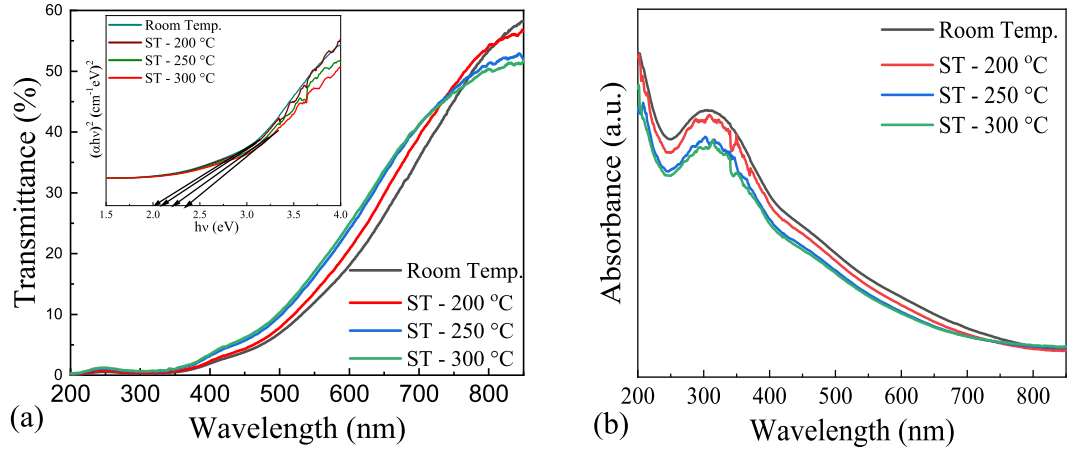}
 \caption{(a) Transmittance spectra (with Tauc plot as inset) and (b) absorbance spectra, for the copper oxides films deposited at different substrate temperatures.}
 \label{fgr:fig4}
\end{figure*}

Optical transmittance spectra were recorded to study the bandgap variation of the copper oxide thin films and the plots are summarized in Figure \ref{fgr:fig4}. From the absorption spectra, it is clear that the thin films have a critical absorption edge around 300 nm. The optical bandgap can be determined by applying the Tauc model (the plots between ($\alpha$h$\nu$)$^{2}$ versus h$\nu$) \cite{zheng2018phase,shukor2020electrical,hsu2018amorphous,hashim2015electrical} for the films and are illustrated in the inset. The analysis indicates that the optical gap value increases with deposition temperature with values of 2.02, 2.09, 2.21, and 2.35 eV for room temperature, 200, 250, and 300 \textdegree C deposition respectively. The reason could be the partial filling of the conduction band, which results in blocking of the lowest states causing a widening of the optical band gap (blue shift) known as the Burstein-Moss shift i.e., shifting of the unoccupied states towards higher energies in the allowed bands.

\subsection{Electrical Properties}
The electrical properties were studied using the standard four-probe method. The sheet resistance as a function of different substrate temperatures are compiled in Table \ref{tbl:table2}. The conductivity($\sigma$) of the samples were calculated using the following formula

\begin{equation}
    \rho = R_s\times t
\end{equation}
\begin{equation}
    \sigma = \frac{1}{\rho}
\end{equation}
where $R_s$ is the sheet resistance and $t$ is the thickness of the copper oxide film.
\begin{table*}
\small
  \caption{The variations in the electrical properties of the copper oxides films deposited at different substrate temperatures}
  \label{tbl:table2}
  \begin{tabular*}{\textwidth}{@{\extracolsep{\fill}}llll}
    \hline
 Substrate Temperature (\textdegree C) & Sheet Resistance (M$\Omega$/sq.) & Conductivity (Sm$^{-1}$) & Band Gap (eV)\\\hline
    Room Temperature & 0.02 & 646.6 & 2.02\\
    200 & 0.19 & 78.6 & 2.13 \\
    250 & 0.07 & 212.0 & 2.21\\
    300 & 0.02 & 711.0 & 2.35\\
    \hline
  \end{tabular*}
\end{table*}

The as-deposited film shows a comparatively lower resistance than the substrate treated film, which may be due to the presence defects and multiple phases of the oxide. The conductivity of the film deposited at the substrate temperature of 200 \textdegree C is the lowest, while conductivity values are higher for the films grown at 250 and 300 \textdegree C. The increase in conductivity might be caused by an increase in the free path of the carrier concentration (higher mobility) or an increase in carrier concentration \cite{bose2005electrical,raship2017effect}. The increase in the conductivity can also be correlated with the growth in the grain size with increasing substrate temperature.

\section{Conclusions}
Thin films of copper oxide have been deposited successfully onto quartz substrates at room temperature, 200, 250, and 300 \textdegree C respectively. XRD analysis revealed that the prepared films at room temperature were polycrystalline in nature with Cu$_2$O as the dominant phase with the preferred orientation along the (110) plane. Structural parameters such as grain size and lattice strain were estimated and found to be strongly dependent on the substrate temperature. Transmittance measurements were used to measure the optical gap which was found to increase with deposition temperature. This highly stable Cu$_2$O thin film can be a promising material for Cu$_2$O based solar cells, anti-reflection coatings due to their increased bandgap and excellent electrical properties. Hence these parameters provide a promising way to control the properties of copper oxide film for desired device applications. Future work will focus on using these films for electrochromic studies.

\section*{Author Contributions}
All authors contributed to the conception and design. The experiments were performed the experiments and characterized by Aarju Mathew Koshy. Dr. A. Sudha helped in analysis of the results and writing the manuscript drafts. The overall work and manuscript preparation was carried out under the supervision of Dr. Satyesh Kumar Yadav and Dr. Parasuraman Swaminathan.

\section*{Conflicts of interest}
 There are no conflicts to declare.

\section*{Acknowledgements}
The work was supported by IIT Madras under the Institute of Eminence project number SB/2021/0850\linebreak/MM/MHRD/008275. The Raman Spectroscopy was carried out at the Material Science Research Center, IIT Madras. The UV-Vis measurements were performed at the Dept. of Physics, IIT Madras. The XRD measurements were recorded at the Common Instruments Facility, ICSR, IIT Madras. The optical profilometry measurements were done at the Centre for NEMS and Nanophotonics (CNNP), Department of Electrical Engineering, IIT Madras. Dr. A. Sudha would like to acknowledge IIT Madras Institute post-doctoral fellowship program for the funding.

%\nolinenumbers

%This is where your bibliography is generated. Make sure that your .bib file is actually called library.bib
\bibliography{library}

\begin{thebibliography}{10}

\bibitem{wong2016current}
Terence~KS Wong, Siarhei Zhuk, Saeid Masudy-Panah, and Goutam~K Dalapati.
\newblock Current status and future prospects of copper oxide heterojunction
  solar cells.
\newblock {\em Materials}, 9(4):271, 2016.

\bibitem{richardson2001electrochromism}
Thomas~J Richardson, Jonathan~L Slack, and Michael~D Rubin.
\newblock Electrochromism in copper oxide thin films.
\newblock {\em Electrochimica Acta}, 46(13-14):2281--2284, 2001.

\bibitem{nolan2006p}
Michael Nolan and Simon~D Elliott.
\newblock The p-type conduction mechanism in cu 2 o: a first principles study.
\newblock {\em Physical Chemistry Chemical Physics}, 8(45):5350--5358, 2006.

\bibitem{kunti2016comparative}
Arup~Kumar Kunti, Shailendra~Kumar Sharma, and Mukul Gupta.
\newblock A comparative study on structural growth of copper oxide deposited by
  dc-ms and hipims.
\newblock {\em ECS Journal of Solid State Science and Technology}, 5(10):P627,
  2016.

\bibitem{rydosz2018use}
Artur Rydosz.
\newblock The use of copper oxide thin films in gas-sensing applications.
\newblock {\em Coatings}, 8(12):425, 2018.

\bibitem{aktar2020solution}
Asma Aktar, Shamim Ahmmed, Jaker Hossain, and Abu Bakar~Md Ismail.
\newblock Solution-processed synthesis of copper oxide (cu x o) thin films for
  efficient photocatalytic solar water splitting.
\newblock {\em ACS omega}, 5(39):25125--25134, 2020.

\bibitem{debbichi2012vibrational}
L~Debbichi, MC~Marco~de Lucas, JF~Pierson, and P~Kruger.
\newblock Vibrational properties of cuo and cu4o3 from first-principles
  calculations, and raman and infrared spectroscopy.
\newblock {\em The Journal of Physical Chemistry C}, 116(18):10232--10237,
  2012.

\bibitem{murali2018thermal}
Dhanya~S Murali and Subrahmanyam Aryasomayajula.
\newblock Thermal conversion of cu4o3 into cuo and cu2o and the electrical
  properties of magnetron sputtered cu4o3 thin films.
\newblock {\em Applied Physics A}, 124(3):1--7, 2018.

\bibitem{zheng2018phase}
Weifeng Zheng, Yue Chen, Xihong Peng, Kehua Zhong, Yingbin Lin, and Zhigao
  Huang.
\newblock The phase evolution and physical properties of binary copper oxide
  thin films prepared by reactive magnetron sputtering.
\newblock {\em Materials}, 11(7):1253, 2018.

\bibitem{winkler2018solution}
Nina Winkler, Stefan Edinger, Jatinder Kaur, Rachmat~Adhi Wibowo, Wolfgang
  Kautek, and Theodoros Dimopoulos.
\newblock Solution-processed all-oxide solar cell based on electrodeposited
  cu2o and znmgo by spray pyrolysis.
\newblock {\em Journal of materials science}, 53(17):12231--12243, 2018.

\bibitem{johan2011annealing}
Mohd~Rafie Johan, Mohd Shahadan~Mohd Suan, Nor~Liza Hawari, and Hee~Ay Ching.
\newblock Annealing effects on the properties of copper oxide thin films
  prepared by chemical deposition.
\newblock {\em Int. J. Electrochem. Sci}, 6(12):6094--6104, 2011.

\bibitem{bose2005electrical}
Anindita Bose, Soumen Basu, Sourish Banerjee, and Dipankar Chakravorty.
\newblock Electrical properties of compacted assembly of copper oxide
  nanoparticles.
\newblock {\em Journal of applied physics}, 98(7):074307, 2005.

\bibitem{papadimitropoulos2005deposition}
G~Papadimitropoulos, N~Vourdas, V~Em Vamvakas, and D~Davazoglou.
\newblock Deposition and characterization of copper oxide thin films.
\newblock In {\em journal of physics: conference series}, volume~10, page 045.
  IOP Publishing, 2005.

\bibitem{paredes2018ultrasonic}
C~Paredes-S{\'a}nchez, RI~S{\'a}nchez-Alarc{\'o}n, O~Hern{\'a}ndez-Silva,
  L~Lartundo-Rojas, G~Alarc{\'o}n-Flores, E~P{\'e}rez-Cappe,
  Y~Mosqueda-Laffita, G~Mesa-P{\'e}rez, C~Falcony, IA~Gardu{\~n}o-Wilches,
  et~al.
\newblock Ultrasonic spray pyrolyzed copper oxide and copper-aluminum oxide
  thin films: optical, structural and electronic properties.
\newblock {\em Materials Research Express}, 6(2):026424, 2018.

\bibitem{chen2016effect}
Jieyi Chen, Honglie Shen, Zihao Zhai, Jinze Li, Wei Wang, Huirong Shang, and
  Yufang Li.
\newblock Effect of substrate temperature and post-annealing on the properties
  of cigs thin films deposited using e-beam evaporation.
\newblock {\em Journal of Physics D: Applied Physics}, 49(49):495601, 2016.

\bibitem{al2019effect}
Naoual Al~Armouzi, Ghizlane El~Hallani, Ahmed Liba, Mustapha Zekraoui, Hikmat~S
  Hilal, Noureedine Kouider, and Mustapha Mabrouki.
\newblock Effect of annealing temperature on physical characteristics of cuo
  films deposited by sol-gel spin coating.
\newblock {\em Materials Research Express}, 6(11):116405, 2019.

\bibitem{babar2015surface}
Shaista Babar, Elham Mohimi, Brian Trinh, Gregory~S Girolami, and John~R
  Abelson.
\newblock Surface-selective chemical vapor deposition of copper films through
  the use of a molecular inhibitor.
\newblock {\em ECS Journal of Solid State Science and Technology}, 4(7):N60,
  2015.

\bibitem{raship2017effect}
NA~Raship, MZ~Sahdan, F~Adriyanto, MF~Nurfazliana, and AS~Bakri.
\newblock Effect of annealing temperature on the properties of copper oxide
  films prepared by dip coating technique.
\newblock In {\em AIP Conference Proceedings}, volume 1788, page 030121. AIP
  Publishing LLC, 2017.

\bibitem{low2013morphology}
Jia~Wei Low, Nafarizal Nayan, Mohd~Zainizan Sahdan, Mohd~Khairul Ahmad, Ali
  Yeon~Md Shakaff, Ammar Zakaria, and Ahmad Faizal~Mohd Zain.
\newblock Morphology, topography and thickness of copper oxide thin films
  deposited using magnetron sputtering technique.
\newblock In {\em RSM 2013 IEEE Regional Symposium on Micro and
  Nanoelectronics}, pages 352--355. IEEE, 2013.

\bibitem{zhang2015cuo}
Suoying Zhang, Hong Liu, Chencheng Sun, Pengfei Liu, Licheng Li, Zhuhong Yang,
  Xin Feng, Fengwei Huo, and Xiaohua Lu.
\newblock Cuo/cu 2 o porous composites: shape and composition controllable
  fabrication inherited from metal organic frameworks and further application
  in co oxidation.
\newblock {\em Journal of Materials Chemistry A}, 3(10):5294--5298, 2015.

\bibitem{wang1996electronic}
Lai-Sheng Wang, Hongbin Wu, Sunil~R Desai, and Liang Lou.
\newblock Electronic structure of small copper oxide clusters: From cu 2 o to
  cu 2 o 4.
\newblock {\em Physical Review B}, 53(12):8028, 1996.

\bibitem{hasabeldaim2017effect}
E~Hasabeldaim, OM~Ntwaeaborwa, RE~Kroon, E~Coetsee, and HC~Swart.
\newblock Effect of substrate temperature and post annealing temperature on
  zno: Zn pld thin film properties.
\newblock {\em Optical Materials}, 74:139--149, 2017.

\bibitem{naveena2019comparative}
D~Naveena, T~Logu, R~Dhanabal, K~Sethuraman, and A~Chandra Bose.
\newblock Comparative study of effective photoabsorber cuo thin films prepared
  via different precursors using chemical spray pyrolysis for solar cell
  application.
\newblock {\em Journal of Materials Science: Materials in Electronics},
  30(1):561--572, 2019.

\bibitem{solache2008raman}
H~Solache-Carranco, G~Juarez-Diaz, M~Galvan-Arellano, J~Martinez-Juarez,
  R~Pena-Sierra, et~al.
\newblock Raman scattering and photoluminescence studies on cu 2 o.
\newblock In {\em 2008 5th International Conference on Electrical Engineering,
  Computing Science and Automatic Control}, pages 421--424. IEEE, 2008.

\bibitem{mary2019sugarcane}
AP~Angeline Mary, A~Thaminum Ansari, and R~Subramanian.
\newblock Sugarcane juice mediated synthesis of copper oxide nanoparticles,
  characterization and their antibacterial activity.
\newblock {\em Journal of King Saud University-Science}, 31(4):1103--1114,
  2019.

\bibitem{chand2014structural}
Prakash Chand, Anurag Gaur, and Ashavani Kumar.
\newblock Structural and optical studies of cuo nanostructures.
\newblock In {\em AIP Conference Proceedings}, volume 1591, pages 262--264.
  American Institute of Physics, 2014.

\bibitem{akgul2014influence}
Funda~Aksoy Akgul, Guvenc Akgul, Nurcan Yildirim, Husnu~Emrah Unalan, and Rasit
  Turan.
\newblock Influence of thermal annealing on microstructural, morphological,
  optical properties and surface electronic structure of copper oxide thin
  films.
\newblock {\em Materials Chemistry and Physics}, 147(3):987--995, 2014.

\bibitem{gommes2019ostwald}
Cedric~J Gommes.
\newblock Ostwald ripening of confined nanoparticles: chemomechanical coupling
  in nanopores.
\newblock {\em Nanoscale}, 11(15):7386--7393, 2019.

\bibitem{shukor2020electrical}
Anmar~H Shukor, Haider~A Alhattab, and Ichiro Takano.
\newblock Electrical and optical properties of copper oxide thin films prepared
  by dc magnetron sputtering.
\newblock {\em Journal of Vacuum Science \& Technology B, Nanotechnology and
  Microelectronics: Materials, Processing, Measurement, and Phenomena},
  38(1):012803, 2020.

\bibitem{hsu2018amorphous}
Ming-Hung Hsu, Sheng-Po Chang, Shoou-Jinn Chang, Wei-Ting Wu, and Jyun-Yi Li.
\newblock Amorphous indium titanium zinc oxide thin film transistor and impact
  of gate dielectrics on its photo-electrical properties.
\newblock {\em ECS Journal of Solid State Science and Technology}, 7(7):Q3049,
  2018.

\bibitem{hashim2015electrical}
H~Hashim, SS~Shariffudin, PSM Saad, and HAM Ridah.
\newblock Electrical and optical properties of copper oxide thin films by
  sol-gel technique.
\newblock In {\em IOP Conference Series: Materials Science and Engineering},
  volume~99, page 012032. IOP Publishing, 2015.

\end{thebibliography}

%This defines the bibliographies style. Search online for a list of available styles.
%\bibliographystyle{unsrt}
\bibliographystyle{hunsrt}
\end{document}